\documentstyle[aps,preprint,prb,prbbib]{revtex}
\tighten
\newcommand{\be}{\begin{equation}}
\newcommand{\ee}{\end{equation}}
\newcommand{\bea}{\begin{eqnarray}}
\newcommand{\eea}{\end{eqnarray}}
\newcommand{\bd}{\begin{displaymath}}
\newcommand{\ed}{\end{displaymath}}
\newcommand{\bi}{\begin{itemize}}
\newcommand{\ei}{\end{itemize}}
\newcommand{\bc}{\begin{center}}
\newcommand{\ec}{\end{center}}
\newcommand{\bfl}{\begin{flushleft}}
\newcommand{\efl}{\end{flushleft}}
\newcommand{\bfr}{\begin{flushright}}
\newcommand{\efr}{\end{flushright}}
\newcommand{\f}{\frac}

\def\6{\partial}  
 \def\d{\delta} \def\ve{\varepsilon}

 \def\ss{\sigma} 

  \def\D{\Delta}

\def\={\!\!\!&=&\!\!\!}
\def\+{\!\!\!&&\!\!\!+~}
\def\-{\!\!\!&&\!\!\!-~}

\begin{document}
\title{.\\
Flow-Equations for the Model of Hybridized Bosons and Fermions}
\author{C.P.Moca}
\address{Department of Physics\\University of Oradea, 3700 Oradea Romania}
\author{I.Tifrea and M.Crisan}
\address{Department of Theoretical Physics\\
University of Cluj, 3400 Cluj, Romania}
\maketitle

\begin{abstract}
Using the flow-equations method we showed analytically the occurrence of 
dispersion for the local bosons in a model of hybridized of local bosons 
and fermions.
\end{abstract}

The boson-fermion model has been 
proposed as a possible scenario for high $T_c$ 
superconductivity \cite{1} and it consists of a mixture of itinerant 
electrons interacting with local bosons. The model proposed initially 
\cite{2} 
for the polaronic superconductivity was recently reconsidered by the authors 
\cite{3} using the Boson-Fermion Hamiltonian
\be
H=(zt-\mu)\sum_{i,\sigma}c_{\sigma}^\dagger c_{i\sigma}-
t\sum_{i,j,\ss}c_{i\ss}^\dagger c_{j\ss}+(\D_B-2\mu)\sum_{i}b_i^\dagger b_i+
V\sum_i(b_i^\dagger c_{i\downarrow}c_{i\uparrow}+h.c.)
\label{e1}
\ee
for a square lattice. In Eq.\ref{e1} $t$ denotes the electron hopping energy, 
$z$ the number of nearest neighbors, $\D_B$ the atomic level of bosons (in 
Ref. \onlinecite{3} these are considered as bi-polarons \cite{4}) and $V$ is 
the boson-fermion coupling constant. The last term of the Hamiltonian 
(\ref{e1}) describes the spontaneous decay and recombination process (See Ref. 
\onlinecite{1}) between bosons and fermions on the same site.

Using this model it was predicted that the breakdown of the Fermi liquid (FL) 
behavior is given by the flattening of the fermionic dispersion as the 
wave vector of the Fermi excitations satisfies the condition $k\simeq k_F$ 
($k_F$ is the Fermi wave vector). A similar model based on the interaction 
of a Fermi system with a bosonic critical mode has been proposed \cite{5} 
using a Kondo-Hubbard Hamiltonian.

The breakdown of the Fermi liquid behavior is found to go hand in hand with 
the occurrence of well defined itinerant bosonic excitations with the energy 
$\D_B^R(q)\sim q^2$. This result obtained in \cite{3} solving  
numerically the self-consistent equations for the fermionic and bosonic 
self-energies will be obtained using the flow-equations of the RNG in 
the recent version proposed by Wegner\cite{6}.

The method proposed in Ref. \onlinecite{6} consists in obtaining differential 
equations for the parameters of the Hamiltonian (\ref{e1}) using the equation
\be
\f{dH(l)}{dl}=\left[\eta(l),H(l)\right]
\label{e2}
\ee
where the transform $\eta(l)$ ($l$ is the flow parameter) is given by
\be
\eta(l)=\left[H_{int},H_0\right]
\label{e3}
\ee
where $H_0$ is the non-interacting part of the Hamiltonian (\ref{e1}) written as
\be
H_0=\sum_{k\ss}\varepsilon(k)c^\dagger_{k\ss}c_{k\ss}+
\sum_{q}\D_Bb_{q}^\dagger b_{q}
\label{e4}
\ee
and $H_{int}$ describes the electron-boson hybridization
\be
H_{int}=\sum_{k}(Vc_{k\uparrow}^\dagger c_{q-k\downarrow}b_q+h.c)
\label{e5}
\ee
From the Eqs. (\ref{e2}-\ref{e5}) we obtain the general flow-equations
\be
\f{dV}{dl}=-V(\ve({\bf k})+\ve({\bf q}-{\bf k})-\D_B)^2
\label{e6}
\ee
\be
\f{d\D_B}{dl}=-2\sum_{{\bf k}}|V|^2(\ve({\bf k})+\ve({\bf q}-{\bf k})-\D_B)
(1-n_{{\bf k}}^\uparrow-n_{{\bf q}-{\bf k}}^\downarrow)
\label{e7}
\ee
\be
\f{d\ve({\bf k})}{dl}=2|V|^2\sum_{{\bf k},\ss}
(\ve({\bf k})+\ve({\bf q}-{\bf k})-\D_B)n_{{\bf q}-{\bf k}}^\ss
\label{e8}
\ee
In the following we are interested in the Eqs. (\ref{e6}) and (\ref{e7}) which 
will be used to calculate the energy of the bosonic excitations. Neglecting 
the spin polarization of the conduction electrons the Eq. (\ref{e7}) becomes
\be
\f{d\D_B}{dl}=-\sum_{{\bf k}}|V|^2(\ve({\bf k})+\ve({\bf q}-{\bf k})-\D_B)
\left[\tanh{\f{\ve({\bf k})}{2T}}+\tanh{\f{\ve({\bf q}-{\bf k})}{2T}}\right]
\label{e10}
\ee
In order to calculate the energy of the bosonic excitations we introduce
\be
J(\ve,l)=|V|^2\sum_{{\bf k}}\d(\ve-\ve({\bf k}))
\label{e11}
\ee
which satisfies the equation
\be
\f{\6 J(\ve,l)}{\6 l}=\sum_{{\bf k}}2V\f{\6 V}{\6 l}\d(\ve-\ve({\bf k}))=
-2\sum_{{\bf k}}|V|^2(\ve({\bf k})-\ve({\bf q}-{\bf k})-\D_B)^2
\d(\ve-\ve({\bf k}))
\label{e12}
\ee
In the limit $q\rightarrow 0$ we approximate this equation by
\be
\f{\6 J(\ve,l)}{\6 l}\simeq -2\sum_{{\bf k}}|V|^2(2\ve({\bf k})-\D_B)^2
\d(\ve-\ve({\bf k}))=-2J(\ve,l)(2\ve-\D_B)^2
\label{e13}
\ee
In order to calculate the $q$-dependence of the $\D_B$ from Eq. (\ref{e10}) 
we approximate
\be
\tanh{\f{\ve({\bf k})}{2T}}+\tanh{\f{\ve({\bf q}-{\bf k})}{2T}}\simeq
2\tanh{\f{\ve({\bf k})}{2T}}\left[1+\f{1}{\sinh{(\ve({\bf k})/T)}}
\f{1}{2T^2}\left(\ve({\bf q})-\f{{\bf k}{\bf q}}{m}\right)\right]
\label{e14}
\ee

The Eq. (\ref{e10}) will be written using Eqs. (\ref{e13}) and (\ref{e14}) as
\be
\f{d\D_B}{d l}\simeq -2\sum_{{\bf k}}|V|^2(2\ve_{{\bf k}}-\D_B)
\tanh{\f{\ve({\bf k})}{2T}}\left(1+\f{1}{\sinh{(\ve({\bf k})/T)}}
\f{\ve({\bf q})-{\bf q}{\bf k}/m}{2T^2}\right)
\label{e15}
\ee
which can be written as
\bea
\f{d\D_B}{dl}&=&-2\int d\ve \d(\ve-\ve({\bf k}))\sum_{{\bf k}}|V|^2
(2\ve({\bf k})-\D_B)\tanh{\f{\ve({\bf k})}{2T}}
\left(1
+\f{\ve({\bf q})-{\bf q}{\bf k}/m}{2T^2\sinh{(\ve({\bf k})/T)}}\right)
\nonumber\\
&=&\int d\ve \f{\6 J(\ve,l)}{\6 l}\f{1}{2\ve-\D_B}\tanh{\f{\ve}{2T}}
\left(1+\f{\ve({\bf q})}{2T^2\sinh{(\ve/T)}}\right)
\label{e16}
\eea
The general solution of Eq. (\ref{e16}) is
\be
\D_B(l)=\D_B^I(0)-\int d\ve \f{J(\ve,0)}{2\ve-\D_B^R}\tanh{\f{\ve}{2T}}
\left(1+\f{\ve({\bf q})}{2T^2\sinh{(\ve/T)}}\right)
\label{e17}
\ee
where $J(\ve,\infty)=0$. In the limit of low temperatures (See Ref. 
\onlinecite{3}) Eq. (\ref{e17}) becomes
\be
\D_B(l)=\D_B^I(0)-\int_{0}^{D} d\ve \f{J(\ve,0)}{2\ve-\D_B^R}
\left[1+\f{\ve({\bf q})}{2T^2}\exp{\left(-\f{\ve}{T}\right)}\right]
\label{e18}
\ee
where $D$ is the band width.
If in this equation we take $\D_B(\infty)=\D_B^R$ and for $J(\ve,0)=1/D$ 
we get for $\D_B^R\gg D$
\be
\D_B^R(q)=\D_B^I+
\f{\ve({\bf q})}{2T^2 D}Ei\left(\f{\D_B^R}{2T}\right)\exp\left[-\f{\D_B^R}
{2T}\right]
\label{e19}
\ee
where $Ei(x)=\int_{-x}^\infty e^{-t}/t dt$, for $x>0$.

Eq. (\ref{e19}) represents the main result of this paper and it can be written 
in the form
\be
\D_B^R(q)=\D_B^I+\f{q^2}{2m^*}
\label{e20}
\ee
where 
\be
m^*=TDm
\label{e21}
\ee
Then we obtained analytically the result predicted numerically in Ref. 
\onlinecite{3}, namely the occurrence of well defined itinerant excitations of 
the localized bosons with a temperature dependent effective mass. 

A quadratic dispersion for the localized bosons was also obtained in Ref. 
\onlinecite{4}, but we have to underline that the result is valid only in 
the long-wave ($q\ll q_c$) and low-energy ($z\ll q v_F$) approximations, the 
exact result containing a term of order $q/q_c$ and a term proportional to 
$q^2$.

The same model as we used has been successfully applied for the study of the 
Anderson impurity model \cite{7},on the calculation of the renormalized 
impurity energy. However, the breakdown of the Landau behavior for the Fermi 
liquid predicted in Ref. \onlinecite{3} is expected if we consider a more 
realistic electron contribution in the starting Hamiltonian. At this stage 
of investigation we mention that this behavior is expected due to the 
analogy between the Hamiltonian (\ref{e1}) and the Anderson Hamiltonian 
with finite range interactions which presents a non-Fermi behavior \cite{8}.
In this approximation we neglected the possibility of condensation of the
bosons\cite{9},but this problem  was considered for charged bosons  by 
Alexandrov\cite{10} taking the  Coulomb screening in the  Cooper pairing.
However, a correct evaluation in\cite{10} of the bosonic  energy showed the
existence of a linear term in the  wave  vector q  and the quadratic term
obtained in  this paper.

\section* { Acknowledgements}
Enlightening discussions with A. E. Ruckenstein  and  the e-mail correspondence
with A. Mielke are greatly appreciated.

\end{document}